\pdfoutput=1

\documentclass[3p,times,twocolumn,procedia]{elsarticle}

\usepackage{ecrc}

\volume{00}

\firstpage{1}

\journalname{Nuclear Physics B Proceedings Supplement}

\runauth{J. Kaspar}

\jid{nuphbp}

\jnltitlelogo{Nuclear Physics B Proceedings Supplement}

\CopyrightLine{2015}{Published by Elsevier Ltd.}

\usepackage{amssymb}

\usepackage[figuresright]{rotating}

\newcommand{\gm}{$(g$\,$-$\,$2)$}
\newcommand{\pb}{PbF$_2$}

\begin{document}

\begin{frontmatter}



\dochead{}

\title{Status of the Fermilab $g\!-\!2$ experiment}


\author{J. Kaspar\\for the Fermilab E989 collaboration\corref{cor1}}
\ead{kaspar@uw.edu}
\address{CENPA and Department of Physics, University of Washington, Seattle, WA 98195}

\begin{abstract}
    The upcoming muon \gm~experiment at Fermilab will measure the anomalous magnetic moment of the muon
    to a relative precision of 140\,ppb, 4 times better than the previous experiment at BNL.
    The new experiment is motivated by the persistent
    3\,--\,4 standard deviations difference between the experimental value and the Standard Model prediction,
    and it will have the statistical sensitivity necessary to either refute the claim or confirm it with a  confidence level exceeding a discovery threshold.
    The experiment is under construction and scheduled to start running in early 2017.
\end{abstract}

\begin{keyword}
g-2 experiment \sep anomalous magnetic moment \sep precision test of Standard Model
\end{keyword}

\end{frontmatter}

\section{Motivation}
\label{jk-g-2-motivation}

The most recent experiment at BNL measured the value of anomalous magnetic dipole moment $a_\mu =  116\,592\,089 (63)\,\times\,10^{11}$ \cite{Bennett:2006fi}; i.e., the experiment achieved the relative precision of 540\,ppb. The experimental value is in more than 3 standard deviations tension with the Standard Model predictions \cite{Hagiwara:2011af}. Despite intense theoretical efforts the discrepancy persists \cite{Benayoun:2014tra}.
The new muon \gm~experiment~\cite{Carey:2009zzb} at Fermilab will improve the precision of the experimental value by a factor of 4.

 \begin{figure}[t]
   \includegraphics[width=1.0\columnwidth]{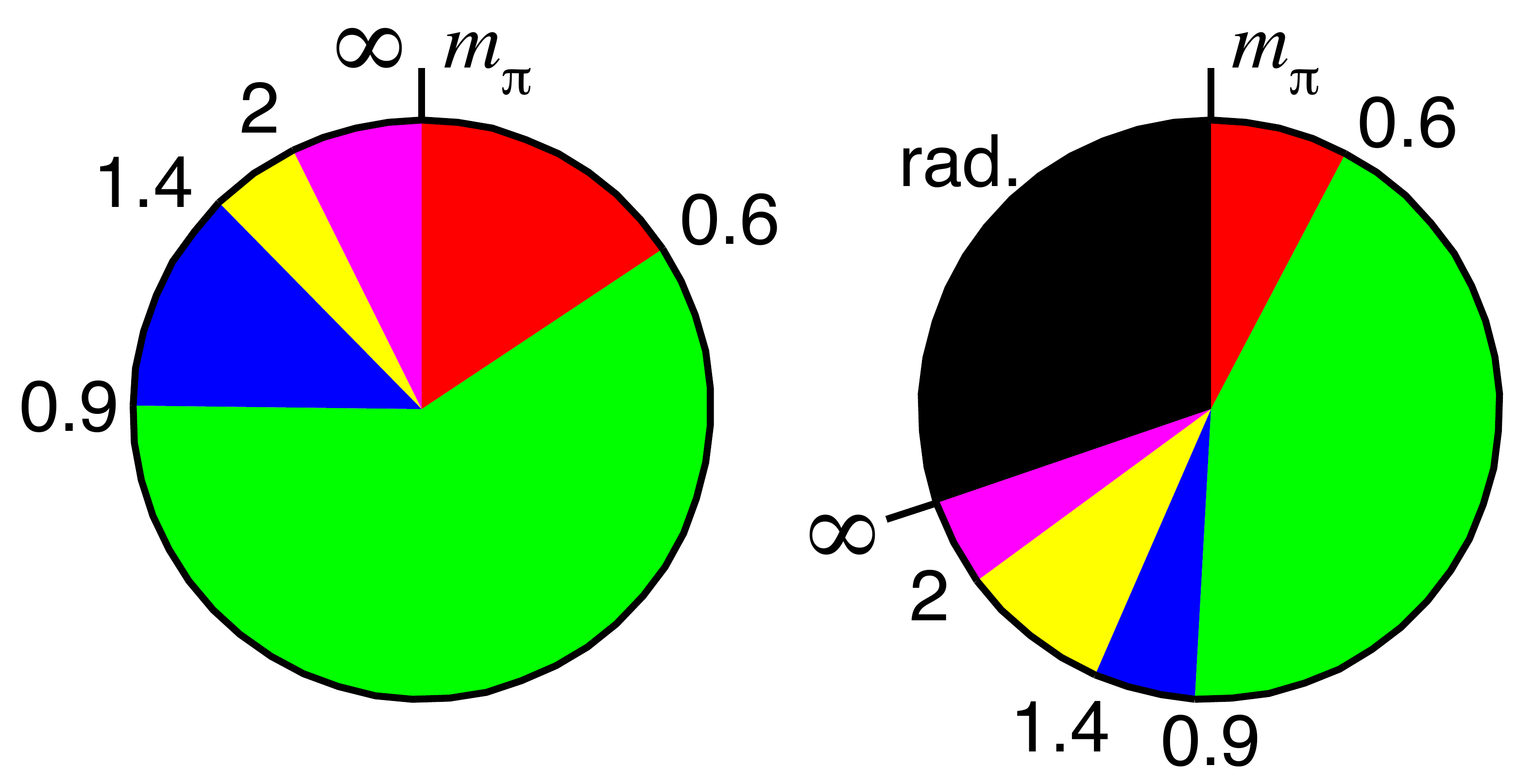}

    \caption{SM contributions to the hadronic vacuum polarization (left) and their variances (right) cited from \cite{Hagiwara:2011af}.}
    \label{fg:gm2-sm-pie}
\end{figure}

The anomalous part of muon magnetic dipole moment is dominated by QED contributions. However, these contributions, together with tiny electro-weak and Higgs terms can be numerically calculated with sufficient precision. The uncertainty of the theoretical value is dominated by hadronic contributions: hadronic vacuum polarization is extracted from electron-positron cross-sections via the optical theorem and a dispersive integral; and light-by-light contributions are mainly calculated using perturbative QCD, although a method utilizing experimental pion form-factors has been proposed recently.

Many electron-positron colliders keep delivering new data sets that improve the theoretical value in two ways: 1. direct improvement from better statistics, and 2. vintage data sets that are not statistically compatible with the latest data can be retired. As the total uncertainties on these cross-sections shrink, the relative contribution from radiative corrections rises (fig.~\ref{fg:gm2-sm-pie} right). These terms are derived from dedicated MC efforts.

 The \gm~experiment is a completeness test of the Standard Model, in the sense that any new physics can manifest itself via loop corrections to the vacuum polarization and make the experimental value of $a_\mu$ deviate from the SM prediction.

\section{Principles of the experiment}
\label{jk-g-2-principles}

Conceptually, \gm~is a very straightforward experiment relying on parity violation in the weak decay to determine the anomalous precession frequency of a muon, $\omega_{\mathrm a}$. Combining $\omega_{\mathrm a}$ with the Larmor frequency $\omega_{\mathrm p}$ of a free proton in the same magnetic field yields $a_\mu$:
\begin{equation}
a_\mu = \frac{\frac{\omega_{\mathrm a}}{\omega_{\mathrm p}}}{\frac{\mu_\mu}{\mu_{\mathrm p}} - \frac{\omega_{\mathrm a}}{\omega_{\mathrm p}}}\ ,
\end{equation}
where $\mu_\mu / \mu_{\mathrm p}$ is the muon-to-proton magnetic moment ratio externally determined from hyperfine splitting in muonium.

The four key ingredients in the experiment are: 1. a source of highly polarized muons; 2. the relative spin precession of the muon spin and momentum is proportional to the anomalous part of the magnetic dipole moment; 3. the magic muon momentum, which cancels contributions from electrostatic fields; and 4. the parity violation in muon decay transfers the average muon spin information to the daughter positron.

\begin{figure}[t]
   \includegraphics[width=1.0\columnwidth]{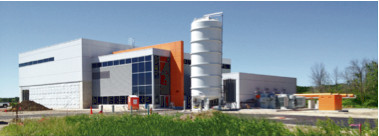}

   \caption{The Muon Campus \gm~building captured in Summer 2014 undergoing construction.}
   \label{fg:MC1-building}
\end{figure}

\subsection{Source of polarized muons}

Pions with the energy of 3.1\,GeV originate from 8.0\,GeV protons hitting an Inconel target. The pions decay in flight into muons with a well defined spin direction due to the parity violation in the weak decay. The decay muons are injected into the magnetic ring to start the spin precession from  known initial spatial and momentum distributions.

\subsection{Anomalous spin precession}

The difference between the cyclotron frequency,
\begin{equation}
\omega_{\mathrm C} = \frac{eB}{mc\gamma} \quad ,
\end{equation}
and the spin precession frequency,
\begin{equation}
\omega_{\mathrm S} = \frac{geB}{2mc} + (1-\gamma)\frac{eB}{\gamma m c} \quad ,
\end{equation}
is proportional to the anomalous magnetic dipole moment,
\begin{equation}
\omega_{\mathrm a} = \omega_{\mathrm S} - \omega_{\mathrm C} = a_\mu \frac{eB}{mc}  \quad ,
\end{equation}
and not the regular full magnetic dipole moment. The experiment measures the muon spin direction as a function of the muon time in the ring after injection.

\subsection{Magic momentum}

Since electrostatic quadrupoles are used for vertical focusing, the muon spin evolution follows:
\begin{equation}
\vec\omega_{\mathrm a} = -\frac{e}{m} \left[ a_\mu \vec{B} - \left( a_\mu - \frac{1}{\gamma^2 - 1} \right) \frac{\vec{\beta} \times \vec{E}}{c}\right]\ .
\end{equation}
A choice of $\gamma=29.3$, corresponding to the muon momentum of 3.094\,GeV/$c$, makes the second term vanish. A very small correction due to momentum spread is handled as a systematic effect.

\subsection{Parity violating muon decay}

After a muon decays, the information on the average muon spin direction is passed on to the daughter positron thanks to parity violation in the weak decay $\mu^+ \rightarrow {\mathrm e}^+ \nu_{\mathrm e}\, \bar{\nu}_\mu$. The positron curls inwards and hits a calorimeter. The hit time and deposited energy are observables in the experiment.

Applying a high energy cut around 1.8\,GeV converts the cosine dependence of muon decay differential cross-section on the polar angle, between the muon spin and positron momentum, into the time domain. Therefore, storing hit times of high-energy positrons in a histogram, across all the muon decays and muon fills, is a practical way to extract the anomalous precession frequency, $\omega_{\mathrm a}$. Correctly disentangling pileup events is critical for the histogram shape to stay statistically compatible with the cosine model.

\section{Where the factor 4 improvement comes from}
\label{jk-g-2-factor4}

The result of the previous BNL experiment was statistically limited. The new Fermilab project will rely on the muon production facility at Fermilab to deliver 21 times more muons into the same  well established magnetic ring, which will be instrumented with new electromagnetic calorimeters and in-vacuum straw tracker detectors. Great attention is being paid to advanced simulation techniques and statistical analysis methods.

The total uncertainty of 140\,ppb breaks equally into statistics and systematics, each contributing  100\,ppb. Further, the systematic error splits between magnetic field and anomalous precession frequency measurement, each accounting for 70\,ppb, and the estimates are based on designed improvements compared to the results achieved at BNL.

\subsection{New accelerator}

The \gm~experiment will be running in the new Muon Campus, which is under construction at Fermilab (fig.~\ref{fg:MC1-building}). Since the previous experiment was statistically limited, the principle goal is to deliver 21x the statistics of the previous effort, which translates into $1.6\,\times\,10^{11}$ recorded daughter positrons above the energy threshold of 1.86\,GeV.

Protons accelerated in the upgraded Linac and Booster ($4\,\times\,10^{12}$ protons per batch) are adiabatically re-bunched in the Recycler and led to the Inconel target. Secondary beam pions then travel around the Delivery Ring (former anti-proton Debuncher Ring), where the pions decay into muons, and finally the beam continues into the muon storage ring. The longer path is beneficial for all the unwanted meson backgrounds to decay away.

Not only will the number of delivered muons be higher, their quality will be much better too: the muon polarization is greater than 0.95; they come with very narrow momentum spread ($<2\,\%$); and the muon beam is practically free of protons or pions, which created a major background at BNL. The higher fill frequency (by about a factor of 3) of the Fermilab machine is one of the major reasons behind the improved muon rate.

\subsection{Magnetic ring}

The 14-m-diameter superconducting magnet generating a highly uniform 1.45\,T field was moved from BNL to Fermilab in 2013; it has since been reconstructed and prepared for commissioning.  The uniformity of the field, averaged over azimuth, must be much better than 1\,ppm, and local field variations are smaller than 100\,ppm. An extended campaign to shim the magnetic field will take place in 2015.

The ring is equipped with weak-focusing electrostatic quadrupoles.
Compared to the previous experiment, higher HV will be applied to the quadrupole electrodes resulting in a higher operating value of the field index (tune), which will decrease systematic errors from coherent betatron oscillation.

\begin{figure}[t]
   \includegraphics[width=1.0\columnwidth]{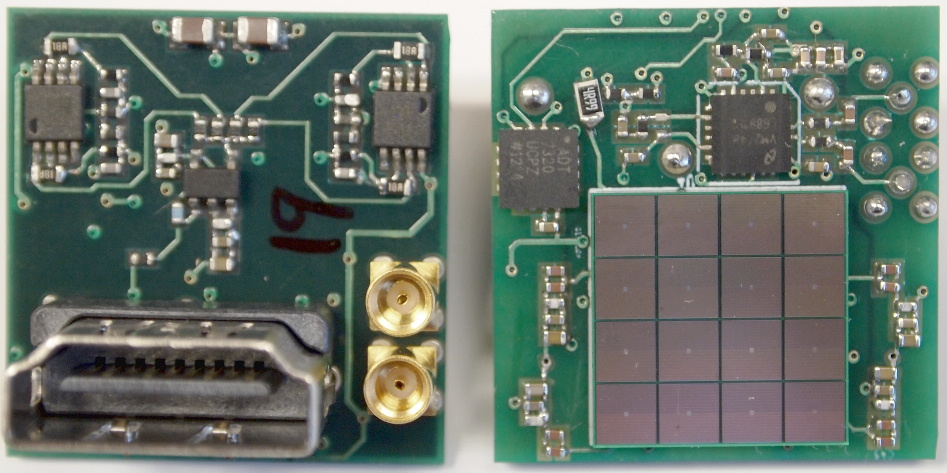}

    \caption{A surface-mounted 16-channel silicon photo-multiplier soldered on an amplifier board
    collects Cherenkov photons in the \gm~calorimeters.}
    \label{fg:SiPM-photo}
\end{figure}

\subsection{Calorimeters}

After a muon decays into a positron and neutrinos, the positron has insufficient energy to remain on the magic orbit in the ring. It curls inward where its energy is detected by \pb~Cherenkov calorimeters which are read out by large area silicon photo-multipliers (SiPM). There are 24 electromagnetic calorimeters sitting on the inside radius of the storage ring to accurately measure the hit times and energies of the positrons.

Systematic uncertainties related to particle pileup played a major role in the previous experiment. The new calorimeter design addresses the pileup challenges by segmenting each calorimeter into 54 independent \pb~crystals, to mitigate spatial pileup, and by using a Cherenkov absorber coupled to fast SiPMs (depicted in Fig.~\ref{fg:SiPM-photo}) with a PMT-like pulse shape (FWHM $<10$\,nsec), to handle particles hitting the same spot on a calorimeter as close in time as 3\,nsec. Non-magnetic SiPMs successfully run in, and at the same time do not disturb, the highly uniform 1.45\,T magnetic field of the storage ring. The charge signal is  recorded continuously during a muon fill by a custom made 12-bit depth 800\,MHz digitizer.

Other calorimeter characteristics, such as energy resolution, light yield, resistance to high rate, and energy linearity, either exceed or meet performance of previous PMT-coupled arrays. A laser calibration system was demonstrated to set and monitor the SiPM gain to a relative precision of better than $10^{-4}$ per hour. A prototype calorimeter successfully passed a test run at SLAC \cite{Fienberg:2014kka} in 2014.

\subsection{Tracking detectors}

Another major group of systematic effects is related to the muon beam profile in the ring and its development during the 700\,$\mu$sec long measuring period.
New tracking detectors will improve the understanding of the muon spatial distribution, momentum spread, and coherent betatron oscillation parameters, and will reduce their contributions to the systematic budget by more than factor 4.
Complementary to the calorimeters, the trackers will verify the pileup separation techniques; and they will measure the upward vs. downward average decay slope, which is sensitive to a possible muon electric dipole moment.

The tracker design relies on 3 tracking stations around the magnet ring, with 9 modules per station. Each module has 4 layers of straws arranged in two doublet planes oriented $\pm7.5^\circ$ from the vertical direction. The straws, with the diameter of 5\,mm, are made from golden coated aluminized Mylar and read out by custom TDC chips.

Similar detailed attention is paid to other auxiliary detectors that will monitor the beam profile of the muons as they are injected into the ring and, on occasions, fiber-harp profile detectors can be rotated into the storage volume to measure the stored muon beam properties directly.
Increased performance and redundancy in all the detectors is the key for the reduced systematic errors to match the improved statistics in the measurement of $\omega_{\mathrm a}$.

\subsection{Magnetic field measurement}

The value of anomalous magnetic moment is derived from $\omega_{\mathrm a}$ and the value of the magnetic field averaged over the ring volume in the azimuthal direction. Although the magnetic field is precisely shimmed to produce the most uniform magnetic field possible, periodical mapping with nuclear magnetic resonance (NMR) probes in exactly the same region where muons propagate is necessary.

Four hundred NMR probes are placed in permanent positions around the ring, several others are carried inside a refurbished trolley running around the ring. The probes are filled by petroleum jelly for improved reliability and decreased temperature dependence.

Finally, the absolute calibration of the magnetic field with respect to the Larmor frequency of the free proton relies on a spherical NMR probe filled with water. He-3 is being investigated as a complementary method for the absolute calibration.

\section{Timeline}
\label{jk-g-2-conclusions}

The experiment completed the DOE CD-2 review in July 2014. The collaboration of 33 institutions from 8 countries reassembled the magnetic ring in the new building, and the ring is ready to undergo cryogenics tests followed by a test ramp cycle.
Extensive magnet shimming will occupy most of 2015, with the accelerator construction running in parallel. First commissioning runs are planned for late 2016. The experiment is scheduled to start running in early 2017, and deliver first results shortly after that.

\begin{figure}[t]
   \includegraphics[width=1.0\columnwidth]{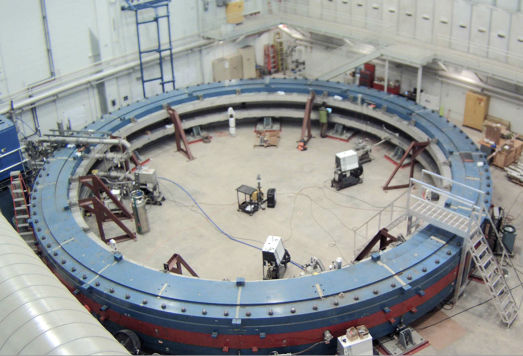}

    \caption{The reassembled magnetic ring captured in January 2015.}
    \label{fg:gm2-ring}
\end{figure}



\section*{Acknowledgements}

This research was supported by the National Science Foundation MRI award PHY-1337542.
This material is based upon work supported by the U.S. Department of Energy Office of Science, Office of Nuclear Physics under Award Number DE-FG02-97ER41020.



\bibliographystyle{elsarticle-num}
\bibliography{gm2}






\end{document}